\documentclass[aps,prb,twocolumn,showpacs,amsmath]{revtex4}
\usepackage{graphicx}
\usepackage{bm}

\begin{document}
\def\Xint#1{\mathchoice
   {\XXint\displaystyle\textstyle{#1}}%
   {\XXint\textstyle\scriptstyle{#1}}%
   {\XXint\scriptstyle\scriptscriptstyle{#1}}%
   {\XXint\scriptscriptstyle\scriptscriptstyle{#1}}%
   \!\int}
\def\XXint#1#2#3{{\setbox0=\hbox{$#1{#2#3}{\int}$}
     \vcenter{\hbox{$#2#3$}}\kern-.5\wd0}}
\def\ddashint{\Xint=}
\def\dashint{\Xint-}

\title{Corbino-geometry Josephson weak links in thin superconducting films}

\author{John R.\ Clem}
\affiliation{%
   Ames Laboratory and Department of Physics and Astronomy, \\
   Iowa State University, Ames, Iowa, 50011--3160}

\date{\today}

\begin{abstract} 
I consider a Corbino-geometry  SNS (superconducting-normal-superconducting) Josephson weak link in a thin superconducting film, in which current enters at the origin, flows outward, passes through an annular Josephson weak link, and leaves radially.  In contrast to sandwich-type annular Josephson junctions, in which the gauge-invariant phase difference obeys the sine-Gordon equation, here the gauge-invariant phase difference obeys an integral equation. I present exact solutions for the gauge-invariant phase difference across the weak link when it contains an integral number $N$ of  Josephson vortices and the current is zero.  I then study the dynamics when a current is applied, and I derive the effective resistance and the viscous drag coefficient; I compare these results with those in sandwich-type junctions.  I also calculate the critical current when there is no Josephson vortex in the weak link but there is a Pearl vortex nearby.  
\end{abstract}


\maketitle

\section{\label{intro}Introduction}

Thin-film annular Josephson weak links have been proposed\cite{Gaitan96,Gaitan01,Plerou01} as a test bed for the observation of the influence of the Berry phase\cite{Berry84} on the dynamics\cite{Ao93} of a vortex trapped in the weak link.
Recent experiments have been carried out by R. H. Hadfield et al.\cite{Hadfield03} in Corbino-geometry thin-film annular Josephson weak links, in which the weak links are in the same plane as the electrodes.  The weak links were fabricated using a focused-ion-beam technique in a superconductor/normal-metal (Nb/Cu) bilayer to mill a 50 nm trench in the superconducting layer to form a weak-link SNS junction.  In the following, I theoretically examine the properties of a thin-film annular Josephson weak link in an idealized Corbino geometry, in which current enters at the origin, flows outward, passes through an annular Josephson weak link, and leaves radially. 

The topological differences between annular weak links and straight weak links of finite length produce striking differences in behavior. For example, since only integral numbers $N$ of flux quanta can be present in annular weak links, their critical currents are zero when $N \ne 0$, whereas arbitrary amounts of flux can enter finite-length weak links, such that their critical currents are usually continous functions of the applied magnetic field.

\begin{figure}
\includegraphics[width=8cm]{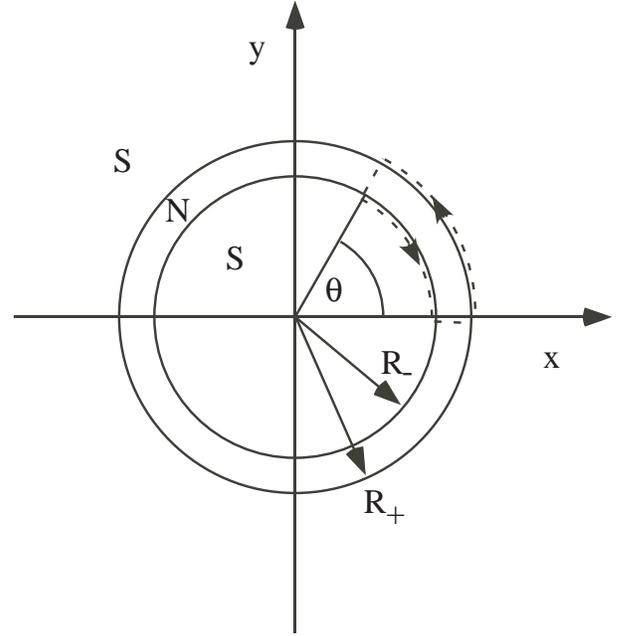}
\caption{%
Sketch of the thin-film  annular Josephson weak link of inner and outer radii $R_- = R - d_N/2$ and $R_+ = R+d_N/2$, considered here.  (The weak-link width $d_N$ is shown greatly exaggerated.) Current enters the film at the origin and leaves radially at a distance well beyond $R_+$. Dashes show the integration contour used to derive Eq.\ (\ref{Deltagammaintegral}).}
\label{annularJJ}
\end{figure}

I consider here only thin films of thickness $d$ less than the London penetration depth $\lambda$, in which the current density $\bm j$ is practically uniform across the thickness, and the characteristic length governing the spatial distribution of the magnetic field distribution is the Pearl length,\cite{Pearl64} 
\begin{equation}
\Lambda = 2 \lambda^2/d.
\label{Pearl}
\end{equation}
Figure \ref{annularJJ} shows the Corbino-geometry SNS Josephson weak link considered.  Current, supplied to the inner superconducting (S) film at the origin, flows radially outward and passes through the annular weak link (N) of inner and outer radii $R_- = R - d_N/2$ and $R_+ = R+d_N/2$, where $d_N \ll R$, and continues to flow radially outward through the outer superconducting (S) film.  

For simplicity, I consider only the case for which $\Lambda \gg R$.  
When a Josephson vortex is trapped in the weak link or a Pearl vortex\cite{Pearl64} is situated in the vicinity of $R$, the magnetic flux $\phi_0 = h/2e$ carried up through the film is spread out over an area of order $\pi \Lambda^2$, so that the corresponding magnetic flux density is very weak.  Although we can neglect the magnetic field generated by the vortex, it is essential to take into account the spatial distribution of the current density $\bm j$ or the sheet-current density $\bm K = \bm j d$. 

In thin-film junctions or weak links\cite{Mints94,Kogan01,Moshe08} there is a second important length scale, which characterizes the spatial variation of the gauge-invariant phase across the junction,\cite{Ivanchenko90,Gurevich92,Ivanchenko95,Kuzovlev97}
\begin{equation}
\ell = \phi_0/4\pi \mu_0 \lambda^2 j_c= \phi_0/2\pi \mu_0 \Lambda K_c,
\label{ell}
\end{equation}
in SI units, where $j_c$ (assumed to be independent of $\theta$ in Fig.\ \ref{annularJJ}) is the maximum Josephson current density that can flow radially as a supercurrent through the weak link and $K_c = j_c d$ is the maximum Josephson sheet-current density.  The case $\ell \gg 2\pi R$ corresponds to the small-junction limit in straight finite-length junctions, and $\ell \ll 2\pi R$ the large-junction limit.\cite{Barone82}

The main goals of this paper are to (a) show that when there are 
$N$ Josephson vortices trapped in the weak link, the critical current $I_c$ is zero for all values of the ratio $\ell/2\pi R$, (b) present exact static solutions for the $\theta$ dependence of the gauge-invariant phase difference for arbitrary $N$ for all values of the ratio $\ell/2\pi R$ when the applied current is zero, (c) examine the dynamics when a current is applied, and (d) show how the critical current density is affected by the presence of a nearby Pearl vortex when there is no Josephson vortex trapped in the weak link.

In Sec.\ \ref{phase}, I derive the basic equation for the gauge-invariant phase difference $\phi(\theta)$ across the weak link and note that there are three additive contributions to $\phi'(\theta) = d\phi(\theta)/d\theta$ to be considered. I examine in Sec.\ \ref{N}  the contribution due to $N$ flux quanta in the weak link, in Sec.\ \ref{P}  the contribution due to a Pearl vortex pinned nearby, and in Sec.\ \ref{J}  the contribution due to Josephson currents.  In  Sec.\ \ref{general}, I derive the integral equations connecting $\phi'$ and $\sin\phi$.   I present exact solutions for the gauge-invariant phase difference in a thin-film annular Josephson weak link containing a single Josephson vortex ($N = 1$) in Sec.\ \ref{exactsolution} and an arbitrary number $N$ of Josephson vortices in Sec.\ \ref{exactallN}, and for all cases I work out some consequences for the vortex dynamics when a net current $I$ is applied. I calculate in Sec.\ \ref{NearbyPearlVortex} the critical current of the annular weak link when there is a Pearl vortex nearby, and I briefly summarize all results in Sec.\ \ref{summary}.  Appendix \ref{Ninslot} contains general expressions for the vector potential and sheet-current density generated by $N$ flux quanta in a narrow circular slot of radius $R$, Appendix \ref{Kappendix} contains details of the Josephson-current-generated sheet current, and Appendix \ref{sandwich} presents a brief comparison with the properties of sandwich-type annular junctions.

\section{\label{phase}Gauge-invariant phase difference}

In the context of the Ginzburg-Landau (GL) theory,\cite{deGennes,StJames69} the superconducting order parameter can be expressed as $\Psi = \Psi_0 f e^{i\gamma}$, where $\Psi_0$ is the magnitude of the order parameter in a uniform sample, $f = |\Psi|/\Psi_0$ is the reduced order parameter, and $\gamma$ is the phase. Let us assume that the induced or applied current densities are so weak that the suppression of the magnitude of the superconducting order parameter  is negligible, such that $f = 1$. For a thin film in which $d < \lambda$ the second GL equation (in SI units) can be expressed as 
\begin{equation}
\bm K = -\frac{2}{\mu_0 \Lambda}(\bm A +\frac{\phi_0}{2\pi}\nabla \gamma),
\label{GL2}
\end{equation} 
where $\bm K = \bm j d$ is the sheet-current density, $\bm A$ is the vector potential, and  $\bm B = \nabla \times \bm A$ is the magnetic induction. 

With a sinusoidal current-phase relation, the Josephson sheet-current density in the radial $\rho$ direction across the weak link is $K_\rho(\theta) = K_c \sin \phi(\theta)$, where $K_c$  is the maximum Josephson sheet-current density and $\phi(\theta)$ is the gauge-invariant phase difference between the inner ($\rho < R_-$)  and outer ($\rho>R_+$) superconducting banks,
\begin{equation}
\phi(\theta) = \gamma(R_-,\theta) - \gamma(R_+,\theta) -\frac{2\pi}{\phi_0}\int_{R_-}^{R_+}
A_\rho(\rho,\theta) d\rho,  
\label{Deltagammadefinition}
\end{equation}
where $R_\pm = R \pm d_N/2$.
A simple relation between $\phi(\theta) $ and the sheet-current densities at $\rho = R_-$ and  $\rho = R_+$ can be obtained by integrating the vector potential $\bm A$ around a loop of width a few coherence lengths larger than $d_N$ enclosing the weak link with one end of the arc at $\theta' = 0$ and the other end  at $\theta' = \theta$, as shown by the dashed contour in Fig.\  \ref{annularJJ}. (The weak link may cause a proximity-induced suppression of the order parameter over a distance of the order of the coherence length $\xi$ into the superconductor.  I assume here that $\xi \ll d_N$.) Since $d_N \ll \Lambda$, we can neglect the magnetic flux up through the loop.  Making use of Eq.\ (\ref{GL2}) for those portions of the integration along the inner and outer boundaries of the weak link, we obtain
\begin{eqnarray}
\phi(\theta)& = &\phi(0) \nonumber \\
&+&\!\!\!\frac{\pi \mu_0 \Lambda}{\phi_0}
\!\!\int_0^\theta\![R_+ K_\theta(R_+,\theta')\!-\!R_- K_\theta(R_-,\theta')]d\theta',
\label{Deltagammaintegral}
\end{eqnarray} 
such that the gauge-invariant phase difference obeys
\begin{equation}
\frac{d\phi}{d\theta} = \frac{\pi \mu_0 \Lambda}{\phi_0} [R_+ K_\theta(R_+,\theta)-R_- K_\theta(R_-,\theta)].
\label{Deltagammaderivative}
\end{equation}
  
When a current $I$ enters at the origin and there is neither a Josephson vortex trapped in the weak link nor a Pearl vortex pinned nearby, the sheet-current $\bm K$ has a radial component $K_\rho = I/2\pi\rho$ but no azimuthal component ($K_\theta = 0$), such that the gauge-invariant phase difference  $\phi$ is independent of $\theta$.  Since $d_N \ll R$, the radial current of the weak link is $I = 2\pi R K_c \sin \phi$ to good approximation, and the maximum supercurrent that can flow without producing a voltage across the weak link is the critical current, $I_{c0} =   2\pi R K_c$.   

On the other hand, when flux quanta are trapped in the weak link or a Pearl vortex is pinned nearby, azimuthal symmetry is destroyed, the radial component of the sheet-current density $K_\rho$ varies as a function of $\theta$, and the azimuthal component of the sheet-current density $K_\theta$ has the property that $[R_+ K_\theta(R_+,\theta)-R_- K_\theta(R_-,\theta)]\ne 0$, such that $\phi$ varies with $\theta$ according to Eq.\ (\ref{Deltagammaderivative}). 
The net supercurrent carried through the weak link is $I = I_{c0}\overline{\sin\phi}$, where the bar denotes the average over $\theta$, and the critical current is given by its maximum value,  $I_c = I_{c0} |\overline{\sin\phi}|_{max}$. 

Although there are nonlinearities associated with the properties of Josephson weak links, it is important to note that Eq.\ (\ref{Deltagammaderivative}) is a linear equation.  Just as the net sheet-current density $\bm K$ can be written as a linear sum of contributions, so also can $d\phi/d\theta$ be written as a linear sum of contributions.
Since we are interested in the behavior when flux quanta are in the annulus, pinned vortices are nearby, and radial Josephson currents flow, we need to calculate the effects of the linear superposition of all three of the corresponding contributions to the sheet-current density $\bm K$ and the $\theta$ derivative of the gauge-invariant phase difference,
\begin{equation}
\frac{d\phi(\theta)}{d\theta} = \Big[\frac{d\phi(\theta)}{d\theta}\Big]_N + \Big[\frac{d\phi(\theta)}{d\theta}\Big]_P+ \Big[\frac{d\phi(\theta)}{d\theta}\Big]_J, 
\end{equation}
each of which can be calculated from the corresponding discontinuity in $\rho K_\theta(\rho,\theta)$ across the annulus at $\rho = R$ [Eq.\ (\ref{Deltagammaderivative})]. We next examine each of these three contributions in turn.

\section{\label{N}$N$ flux quanta in the annular weak link}

Suppose that $N$ flux quanta are trapped in the annular weak link in the absence of any nearby Pearl vortices or any Josephson currents across the junction.  Since we are considering only the case that $\Lambda \gg R$, we can neglect the vector potential term in Eq.\ (\ref{GL2}).  For $\rho < R_-$, the phase $\gamma$ = const and  $\bm K = 0$.   However, for $\rho > R_+$, the phase winds by multiples of $2\pi$.  When there are $N$ flux quanta in the junction, $\gamma = -N\theta$, which generates the azimuthal sheet-current contribution  $K_\theta(\rho) = N\phi_0/\pi \mu_0 \Lambda \rho$ in the region $\rho > R_+$.  From  Eq.\ (\ref{Deltagammaderivative}) we obtain
\begin{equation}
\Big[\frac{d\phi(\theta)}{d\theta}\Big]_N=N.
\label{phiprimeN}
\end{equation}
The above current, phase, and field distributions are equivalent to those produced by $N$ Pearl vortices whose cores are distributed uniformly around the circle of radius $R$,\cite{Kogan01} such that the total magnetic flux carried up through the superconducting film is $N \phi_0$. See Appendix \ref{Ninslot} for details. Choosing an integration contour in the shape of a circular sector of radius $\rho > R$ and central angle $\theta$ instead of the contour shown in Fig.\ \ref{annularJJ}, one can show that Eq.\ (\ref{phiprimeN}) is valid for any value of $\Lambda/R$.

\begin{figure}
\includegraphics[width=8cm]{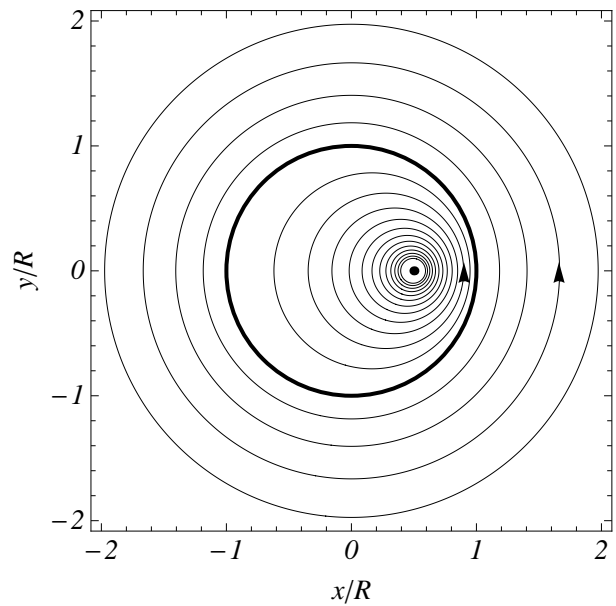}
\caption{%
Streamlines of the sheet current $\bm K_{in}$ generated by a Pearl vortex  at $(x,y) = (\rho_v,0) = (R/2,0)$ (black point) {\it inside} the annulus when $I = 0$ and $N = 0$, obtained as a contour plot of the real part of the complex potential ${\cal G}_{in}$ [Eq.\ (\ref{Gin})].  The bold circle shows the annular weak link, and arrows show the current direction. }
\label{Ginplot}
\end{figure} 
\begin{figure}
\includegraphics[width=8cm]{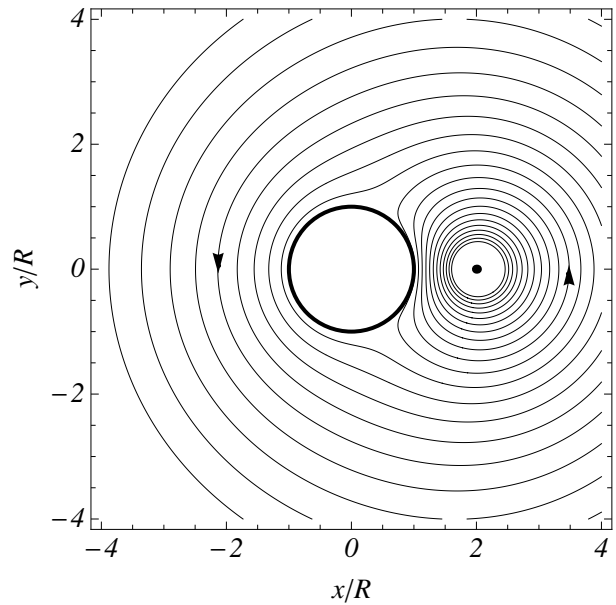}
\caption{%
Streamlines of the sheet current $\bm K_{out}$ generated by a Pearl vortex  at $(x,y) = (\rho_v,0) = (2R,0)$ (black point) {\it outside} the annulus when $I = 0$ and $N = 0$, obtained as a contour plot of the real part of the complex potential ${\cal G}_{out}$ [Eq.\ (\ref{Gout})].  The bold circle shows the annular weak link, and arrows show the current direction.
Note that the length scales differ from those in Fig.\ \ref{Ginplot}.}
\label{Goutplot}
\end{figure}

\section{\label{P}Pearl vortex pinned nearby}

Suppose that a Pearl vortex is pinned at $(x,y) = (\rho_v,0)$ either inside the annular weak link ($\rho_v < R_-$) or outside ($\rho_v > R_+$), but no flux quanta are trapped in the annular slot nor are there any radial Josephson currents across the junction. Since $\nabla \cdot \bm K = 0$ and  
$\nabla \times \bm K = 0$, with the latter equation holding to good approximation because the the magnetic field can be neglected when $R \ll \Lambda$, the method of complex potentials and fields can be used to calculate the sheet-current density generated in response to the Pearl vortex. In general, the complex potential ${\cal G}(\zeta)$ is an analytic function of the complex variable  $\zeta = x + i y$, and the corresponding complex sheet current is 
\begin{equation}
{\cal K}(\zeta) = d{\cal G}(\zeta)/d\zeta =K_y(x,y) + i K_x(x,y).
\end{equation}
Since the radial and azimuthal components of $\bm K$ along the unit vectors $\hat \rho = \hat x \cos \theta + \hat y \sin \theta$ and $\hat \theta = \hat y \cos \theta- \hat x \sin \theta$ are $K_\rho = (H_x x  +H_y y)/\rho$ and $K_\theta = (H_y x - H_x y)/\rho$, we have the relation
\begin{equation}
\tilde {\cal K}(\zeta) = {\cal K}(\zeta)\zeta/\rho = K_\theta(\rho,\theta) + i K_\rho(\rho,\theta),
\label{tildecalK}
\end{equation}
where $x = \rho \cos \theta$, $y = \rho \sin \theta$,  $\rho = \sqrt{x^2+y^2}$, and $\theta = \tan^{-1}(y/x)$. 
          
When $\rho_v < R_-$, the complex potential is 
\begin{eqnarray}
{\cal G}_{in}(\zeta) &=& \frac{\phi_0}{\pi \mu_0 \Lambda} \ln \Big(\frac{\zeta - \rho_v}{\zeta-\rho_i}\Big),\;\rho < R_-, \nonumber \\
&=& \frac{\phi_0}{\pi \mu_0 \Lambda} \ln \zeta, \; \rho > R_+,
\label{Gin}
\end{eqnarray}
where  $\zeta = x + i y$, $x = \rho \cos \theta$, $y = \rho \sin \theta$, and $\rho_i = R^2/\rho_v$ corresponds to the radial coordinate of an image vortex.
Figure \ref{Ginplot} shows a contour plot of the real part of ${\cal G}_{in}$.  The contours correspond to streamlines of $\bm K_{in}$.  

When $\rho_v > R_+$, the complex potential is 
\begin{eqnarray}
{\cal G}_{out}(\zeta) &=& 0,\;\rho < R_-, \nonumber \\
&=&\frac{\phi_0}{\pi \mu_0 \Lambda}[ \ln \Big(\frac{\zeta - \rho_v}{\zeta-\rho_i}\Big) +\ln \zeta], \; \rho > R_+.
\label{Gout}
\end{eqnarray}
Figure \ref{Goutplot} shows a contour plot of the real part of ${\cal G}_{out}$.  The contours correspond to streamlines of $\bm K_{out}$.  

Evaluating ${\cal K}_{in}(\zeta)=d{\cal G}_{in}(\zeta)/d\zeta$ and ${\cal K}_{out}(\zeta)=d{\cal G}_{out}(\zeta)/d\zeta$ at $\rho = R_-$ and $\rho = R_+$ and using Eq.\ (\ref{Deltagammaderivative}), we find that
\begin{equation}
\Big[\frac{d\phi(\theta)}{d\theta}\Big]_P=P(\tilde \rho_v,\theta),
\label{phiprimeP}
\end{equation}
where
\begin{equation}
P(\tilde \rho_v,\theta)=1-\frac{|\tilde \rho_v^2-1|}{\tilde \rho_v^2 +1-2\tilde \rho_v \cos \theta}
\label{Pdefinition}
\end{equation}
with $\tilde \rho_v = \rho_v/R$, 
holds for both $\rho_v < R_-$ and $\rho_v > R_+$.  When $\rho_v = 0$ or $\rho_v = \infty$, $[d\phi(\theta)/d\theta]_P = 0$.  However, as $\rho_v \to R$, $[d\phi(\theta)/d\theta]_P \to 1,$ the same result as Eq.\ (\ref{phiprimeN}) for one flux quantum trapped in the annulus ($N=1$).

\section{\label{J}Josephson currents}

Let us next focus on the contribution to the sheet current $\bm K$ generated by Josephson currents through the junction, ignoring the contributions due to flux quanta in the annular weak link or a nearby Pearl vortex.
To obtain the equation determining how $\phi(\theta)$ varies when the radial Josephson current $K_\rho(R,\theta) = K_c \sin \phi(\theta)$ varies as a function of $\theta$, we start by deriving the Green's function for this problem, assuming that the current $I$ entering at the origin flows through the weak link with a delta-function distribution $K_{0\rho}(R,\theta) = (I/R)\delta(\theta-\theta').$  As in Sec.\ \ref{P}, we can use the method of complex potentials.  The required complex potential is
\begin{equation}
{\cal G}_0(\zeta) = \pm i \frac{I}{2\pi}\ln\Big[\frac{(\zeta-\zeta')^2}{\zeta \zeta'}\Big],
\label{G0}
\end{equation}
where $\zeta = x + i y = \rho e^{i\theta}$, $\zeta' = R e^{i\theta'}$, and the upper (lower) sign holds for $\rho > R$ ($\rho < R$). The corresponding sheet current is 
\begin{equation}
\bm K_0 = \frac{I}{2\pi\rho}\Big[\frac{\hat \rho|\rho^2-R^2| \pm \hat \theta 2\rho R \sin (\theta-\theta') }{\rho^2 + R^2 - 2\rho R \cos (\theta-\theta')}\Big],
\end{equation}
where $\hat \rho = \hat x \cos \theta + \hat y \sin \theta$ and  $\hat \theta= \hat y \cos \theta -\hat x \sin \theta$.  As $\rho \to R$, we obtain $K_{0\rho}(R,\theta) = (I/R)\delta(\theta-\theta')$ and
\begin{equation}
K_{0\theta}(R_\pm,\theta)=\pm \frac{I}{2\pi R_\pm}\cot \Big(\frac{\theta-\theta'}{2}\Big).
\end{equation}

The complex potential for a general distribution of radial Josephson sheet current $K_\rho(\theta) = K_c \sin \phi(\theta)$ can be obtained from Eq.\ (\ref{G0}) by replacing $I$ by $K_c \sin \phi(\theta')R d\theta'$ and integrating over $\theta'$:
\begin{equation}
{\cal G}(\zeta) = \pm i \frac{K_c R }{2\pi} \!\int_{-\pi}^\pi  \sin \phi(\theta')\ln\Big[\frac{(\zeta-\zeta')^2}{\zeta \zeta'}\Big]d\theta'.
\label{G}
\end{equation}
From this expression we find that the radial and azimuthal components of the sheet current associated with the Josephson currents are 
\begin{eqnarray}
K_\rho(\rho,\theta) &=&\!\frac{K_c|\tilde \rho^2\!-\!1|}{2\pi \tilde \rho}\!\!\int_{-\pi}^\pi\!\frac{\sin\phi(\theta')d\theta'}{\tilde \rho^2 +1-2\tilde \rho \cos(\theta\!-\!\theta')}, \label{KrhoJ}\\
K_\theta(\rho,\theta) &=&\pm \frac{K_c}{\pi}\int_{-\pi}^\pi\frac{\sin\phi(\theta')\sin(\theta-\theta')d\theta'}{\tilde \rho^2 +1-2\tilde \rho \cos(\theta\!-\!\theta')},
 \label{KthetaJ}
\end{eqnarray}
where $\tilde \rho = \rho/R$ and the upper (lower) sign holds when $\rho > R_+$ ($\rho < R_-$). 
The terms involving the azimuthal components of the sheet current needed in Eq.\ (\ref{Deltagammaderivative}) are given by the principal-value integral,
\begin{equation}
 R_\pm K_{\theta}(R_\pm,\theta)=\pm \frac{K_c R}{2\pi}\dashint_{-\pi}^\pi  \sin \phi(\theta')\cot \Big(\frac{\theta-\theta'}{2}\Big)d\theta',
\label{Kthetaplusminus}
\end{equation}
such that the Josephson-current contribution to Eq.\ (\ref{Deltagammaderivative}) is
\begin{equation}
\Big[\frac{d\phi(\theta)}{d\theta}\Big]_J = \frac{R}{2\pi \ell}\dashint_{-\pi}^\pi  \sin \phi(\theta')\cot \Big(\frac{\theta-\theta'}{2}\Big)d\theta'.
\label{phiprimeJ}
\end{equation}

\section{\label{general}General equations}

Combining the contributions from Eqs.\ (\ref{phiprimeN}), (\ref{phiprimeP}), and (\ref{phiprimeJ}), we find that that general equation determining the angular dependence of the gauge-invariant phase is
\begin{eqnarray}
\frac{d\phi(\theta)}{d\theta}& =& N + P(\tilde \rho_v,\theta) \nonumber \\
&+& \frac{R}{2\pi \ell}\dashint_{-\pi}^\pi  \sin \phi(\theta')\cot \Big(\frac{\theta-\theta'}{2}\Big)d\theta'.
\label{phiprimegeneral}
\end{eqnarray}
We can invert this  integral equation by making use of 
\begin{eqnarray}
&&\frac{1}{2\pi}\dashint_{-\pi}^\pi  \cot \Big(\frac{\theta'-\theta}{2}\Big)\cot \Big(\frac{\theta'-\theta''}{2}\Big)d\theta' \nonumber \\
&&=2\pi \delta(\theta-\theta'') -1.
\label{PVintegral}
\end{eqnarray}
The result is
\begin{eqnarray}
&&\sin \phi(\theta)= \overline{\sin \phi} \nonumber \\
&& \!\!\!\!+\frac{\ell}{2\pi R}\dashint_{-\pi}^\pi \!\!\Big[\frac{d\phi(\theta')}{d\theta'}\! -\!P(\tilde \rho_v,\theta')\Big] \cot \Big(\frac{\theta'-\theta}{2}\Big)d\theta',
\label{sinphigeneral}
\end{eqnarray}
where
\begin{equation}
\overline{\sin \phi}=\frac{1}{2\pi}\int_{-\pi}^\pi \sin \phi(\theta)d\theta = \frac{I}{I_{c0}},
\end{equation}
The term involving $N$ drops out of Eq.\ (\ref{sinphigeneral}) because 
\begin{equation}
\dashint_{-\pi}^\pi  \cot \Big(\frac{\theta'-\theta}{2}\Big)d\theta'=0.
\end{equation}
Equation (\ref{sinphigeneral}) can be converted back to Eq.\ (\ref{phiprimegeneral}) with the help of Eq.\ (\ref{PVintegral}) and $\int_{-\pi}^\pi P(\tilde \rho_v,\theta)d\theta = 0$.

\section{\label{exactsolution}$N = 1$}

\subsection{\label{statics} Exact solution for the static case when $I = 0$}

\begin{figure}
\includegraphics[width=8cm]{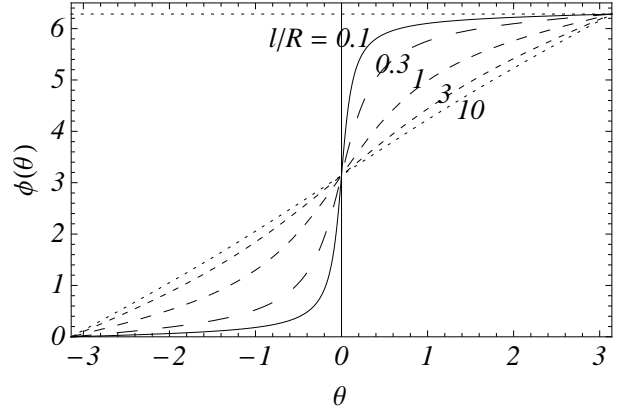}
\caption{%
Exact solution $\phi(\theta)$ [Eq.\ (\ref{phiN=1})] vs $\theta$ for a Josephson vortex ($N = 1$) in a thin-film annular weak link for $\ell/R =$ 0.1, 0.3, 1, 3, and 10.}
\label{phiexactplot}
\end{figure}

When a Josephson vortex is trapped in the annular weak link ($N = 1$) with no Pearl vortex nearby and the current $I$ is zero, the Josephson vortex is stationary, and the gauge-invariant phase obeys
\begin{equation}
\frac{d\phi(\theta)}{d\theta}= 1+ \frac{R}{2\pi \ell}\dashint_{-\pi}^\pi  \sin \phi(\theta')\cot \Big(\frac{\theta-\theta'}{2}\Big)d\theta'.
\label{phiprimeIntEqN=1}
\end{equation}
This equation has an exact solution, corresponding to a Josephson vortex centered at $\theta = 0$,  
\begin{equation}
\phi(\theta)= 2\tan^{-1}\Big[\frac{\tan(\theta/2)}{\tan(\theta_1/2)}\Big]+\pi,
\label{phiN=1}
\end{equation}
where 
\begin{equation}
\tan(\theta_1/2)=\sqrt{1+(R/\ell)^2}-R/\ell,
\label{tantheta1}
\end{equation}
or, alternatively, 
\begin{equation}
\frac{R}{\ell}=\frac{1-\tan^2(\theta_1/2)}{2\tan(\theta_1/2)}.
\label{Rbyell}
\end{equation}
Note that $\phi(-\pi) = 0$, and $\phi(\pi) = 2\pi$; also $\tan(\theta_1/2) \to 1$ when $\ell \to \infty$, and $\theta_1 \to \ell/R \to 0$ when $\ell \to 0.$
Figure \ref{phiexactplot} shows $\phi(\theta)$ vs $\theta$ for a variety of values of $\ell/R$.   

\begin{figure}
\includegraphics[width=8cm]{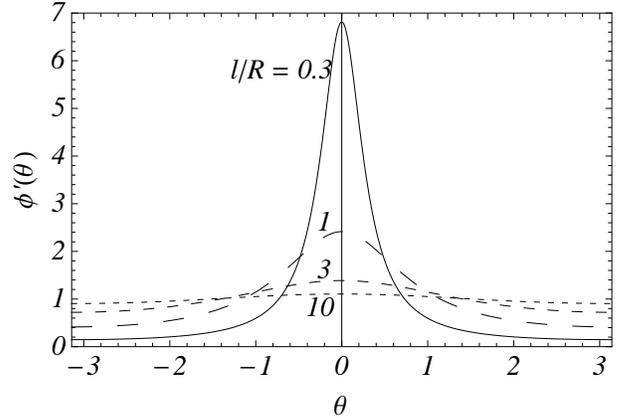}
\caption{%
Plot of the derivative of the exact solution $\phi'(\theta)=d\phi(\theta)/d\theta$ [Eq.\ (\ref{phiprimeN=1})] vs $\theta$ for $N = 1$ and $\ell/R =$ 0.3, 1, 3, and 10.}
\label{phiprimeexactplot}
\end{figure}

Equation (\ref{phiN=1}) yields 
\begin{equation}
\phi'(\theta) = \frac{d\phi(\theta)}{d\theta}= \frac{\tan(\theta_1/2)[1+\tan^2(\theta/2)]}{\tan^2(\theta_1/2)+\tan^2(\theta/2)}.
\label{phiprimeN=1}
\end{equation}
Note that $\phi'(0) = \cot(\theta_1/2) = \sqrt{1+(R/\ell)^2}+R/\ell$ and 
$\phi'(\pm \pi) = \tan(\theta_1/2) = \sqrt{1+(R/\ell)^2}-R/\ell$.
Figure \ref{phiprimeexactplot} shows $\phi'(\theta)$ vs $\theta$ for a variety of values of $\ell/R$.   
\begin{figure}
\includegraphics[width=8cm]{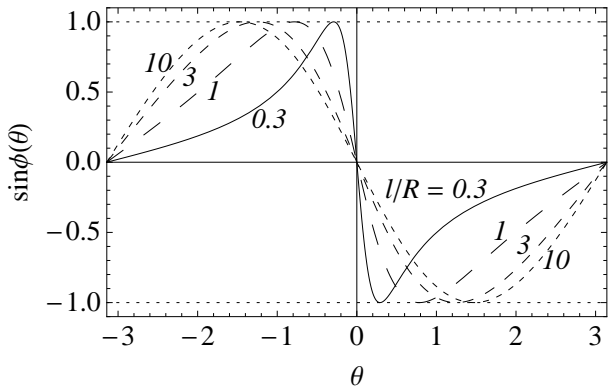}
\caption{%
Plot of the sine of the exact solution $\sin\phi(\theta)$ [Eq.\ (\ref{sinphiN=1})] vs $\theta$ for $N = 1$ and $\ell/R =$ 0.3, 1, 3, and 10.}
\label{sinphiexactplot}
\end{figure}

From Eq.\ (\ref{phiN=1}) we also obtain
\begin{equation}
\sin\phi(\theta)= -\frac{2\tan(\theta_1/2)\tan(\theta/2)}{\tan^2(\theta_1/2)+\tan^2(\theta/2)},
\label{sinphiN=1}
\end{equation}
which has its maximum (+1) and minimum (-1) at $\theta=-\theta_1$ and  $\theta=+\theta_1$.  Defining the angular width $\theta_{core}$ of the Josephson core as the range of $\theta$ values for which $\pi/2 \le \sin\phi(\theta) \le 3\pi/2$, we find for $N = 1$, 
\begin{equation}
\theta_{core} = 2 \theta_1  = 4\tan^{-1}[\sqrt{1+(R/\ell)^2}-R/\ell].
\label{thetacore}
\end{equation}
When $\ell/R \to \infty$, $\theta_{core} = \pi$, and when $\ell/R \ll 1$, $\theta_{core} \approx 2\ell/R.$  See Figs.\ \ref{phiexactplot}, \ref{sinphiexactplot}, and \ref{thetacoreplot}.
Note also that $\sin\phi(\theta) = 0$ at $\theta = -\pi$, 0, and $\pi$.  Since $\sin\phi(\theta)$ is an odd function of $\theta$, $\overline{\sin\phi} = I/I_{c0}=0$. 
Figure \ref{sinphiexactplot} shows $\sin\phi(\theta)$ vs $\theta$ for a variety of values of $\ell/R$.   Figures \ref{phiexactplot}-\ref{sinphiexactplot} and \ref{thetacoreplot} show that the width of the Josephson vortex core increases as $\ell/R$ increases and that as $\ell/R \to \infty,$ the Josephson vortex becomes so spread out that its center can be identified only as the place where $\phi = \pi$ (or an odd multiple of $\pi$). 

\begin{figure}
\includegraphics[width=8cm]{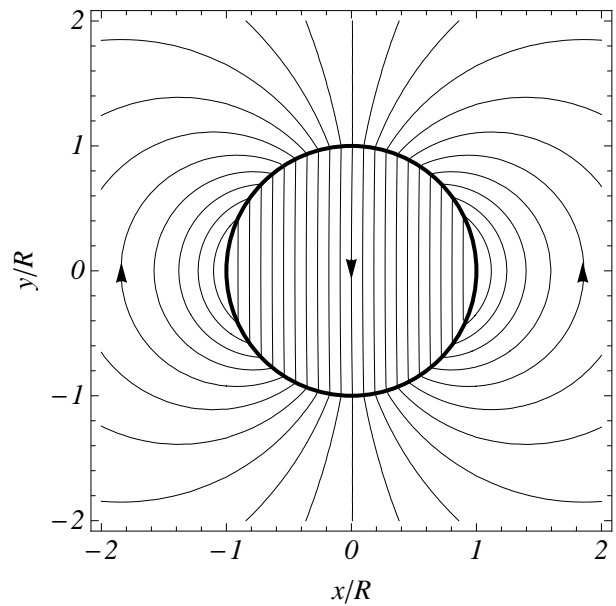}
\caption{%
Dipole-like streamlines of the Josephson-current contribution to the circulating sheet-current density ${\bm K}$ for a single flux quantum $N = 1$ trapped in the weak link when $\ell/R = 30$ and $I=0$, obtained as a contour plot of the real part of the complex Green's function ${\cal G}$ [Eq.\ (\ref{G})] using the exact static solution $\phi(\theta)$ [Eq.\ (\ref{phiN=1})].  The bold circle shows the annular weak link, and the arrows show the current direction. }
\label{G30plot}
\end{figure}

\begin{figure}
\includegraphics[width=8cm]{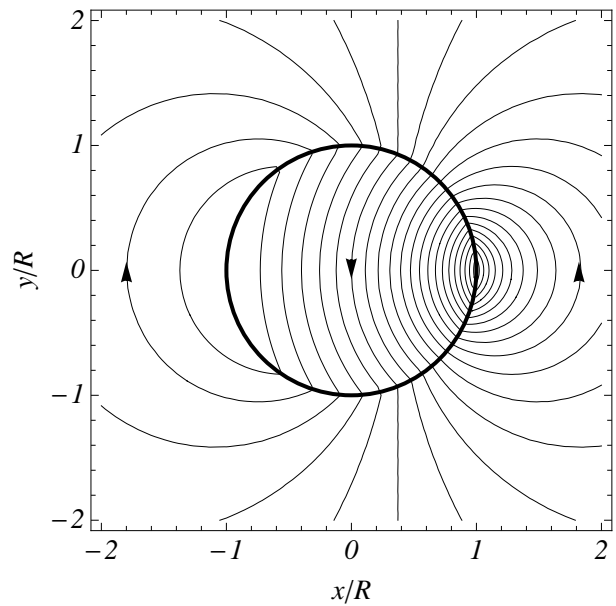}
\caption{%
Streamlines of the Josephson-current contribution to the circulating sheet-current density ${\bm K}$ for a single flux quantum $N = 1$ trapped in the weak link when $\ell/R = 0.3$ and $I=0$, obtained as a contour plot of the real part of the complex Green's function ${\cal G}$ [Eq.\ (\ref{G})] using the exact static solution $\phi(\theta)$ [Eq.\ (\ref{phiN=1})].  The bold circle shows the annular weak link, the Josephson vortex is centered at $(x,y) = (R,0)$, and the arrows show the current direction. }
\label{G0p3plot}
\end{figure}

The sheet-current distribution generated by the Josephson currents can be calculated from Eqs.\ (\ref{KrhoJ}) and (\ref{KthetaJ}) using the exact solution for $\phi(\theta)$ given in Eq.\ (\ref{phiN=1}).  When $\ell/R = \infty$, we have simply $\phi(\theta) = \theta+\pi$, and the sheet current has the simple dipole-like behavior $\bm K = -K_c \hat y$ for $\rho < R_-$ and $\bm K = K_c(R/\rho)^2(\hat y \cos 2\theta-\hat x \sin 2\theta)$ for $\rho > R_+$.  For finite values of $\ell/R$, $\bm K$ can be evaluated analytically as in Appendix \ref{Kappendix} or calculated numerically from Eqs.\ (\ref{KrhoJ}) and (\ref{KthetaJ}).  Streamlines of $\bm K$, generated as contour plots of the real part of $\cal G$ using Eq.\ (\ref{G}), are shown in Figs.\ \ref{G30plot} and \ref{G0p3plot} for $\ell/R = 30$ and $\ell/R = 0.3.$  Recall, however, that the {\it total} sheet-current distribution for $N = 1$ is the sum of the Josephson-current contribution shown here and the contribution $\bm K = \hat \theta \phi_0/\pi \mu_0 \Lambda \rho$ for $\rho > R$, as discussed in Sec.\ \ref{N}.

In sandwich-type annular Josephson junctions [see Appendix \ref{sandwich}], the phase obeys a sine-Gordon equation, which involves the sine and the second derivative of the phase with respect to the coordinate along the junction.\cite{Barone82}  In thin-film annular junctions discussed here, however, the sine of the phase obeys an integral equation, obtained by partial integration of Eq.\ (\ref{sinphigeneral}) with $\overline{\sin\phi}=0$ and $P=0$, 
\begin{equation}
\sin \phi(\theta')= \frac{\ell}{2\pi R}\dashint_{-\pi}^\pi \ln\Big[\csc^2\Big(\frac{\theta'-\theta}{2}\Big)\Big]\phi''(\theta')d\theta',
\label{nosineGordon}
\end{equation}
which is analogous to corresponding integral equations derived in Refs. \onlinecite{Mints94} and \onlinecite{Kogan01}.  The exact solution [Eq.\ (\ref{phiN=1})]  obeys this integral equation, as can be verified by evaluating Eq.\ (\ref{nosineGordon}) with  
\begin{equation}
\phi''(\theta) =  -\frac{\tan(\theta_1/2)[1-\tan^2(\theta_1/2)]\tan(\theta/2)\sec^2(\theta/2)}{[\tan^2(\theta_1/2)+\tan^2(\theta/2)]^2}.
\label{phiprimeprimeN=1}
\end{equation}  

\subsection{\label{dynamics} Dynamical behavior when $I \ne 0$}
To calculate the critical current [see Sec.\ \ref{phase}] of a small Josephson weak link (which corresponds to the case $R/\ell=0$), one usually can start with a non-current-carrying static solution $\phi$ for which $\overline{\sin\phi} = 0$, add a bias phase $\beta$, and then compute $\overline{\sin(\phi+\beta)} = \overline{\cos\phi}\sin\beta$ to conclude that the critical current is proportional to the average $|\overline{\cos\phi}|$.  This procedure remains valid here in the limit $R/\ell = 0$, and the result is $|\overline{\cos\phi}|=|\overline{\cos\theta}|= 0,$ which tells us that the critical current is zero in this case.   However, this procedure fails for finite values of $R/\ell$ because $\phi(\theta) +\beta$ is not a solution of  Eq.\ (\ref{phiprimeIntEqN=1}).  No static current-carrying state can be generated from the exact solution given in Eq.\ (\ref{phiN=1}); there is no solution corresponding to a stationary Josephson vortex in the presence of a current $I$.  In other words,  the critical current $I_c$ of a thin-film annular Josephson weak link is zero for all ratios of $R/\ell$.    

As soon as a current $I$ is applied, the gauge-invariant phase distribution becomes time-dependent and the weak link becomes resistive. The behavior is simplest in the limit $R/\ell = 0$, for which the voltage measured directly across the weak link between $\rho = R_-$ and $R_+$ is, by the Josephson relation, $V = (\hbar/2e)d\phi/dt = I R_n$, where $R_n$ is the normal-state resistance of the annulus.  Since the phase $\phi$ slips by $2\pi$ with a frequency $\nu$, this occurs because the straight-line phase distribution, given by $\phi(\theta) = \theta + \pi$ at time $t = 0$ (similar to the dotted line for $\ell/R =10$ in Fig.\ \ref{phiexactplot}), slides rigidly toward negative values of $\theta$ with an angular velocity $\omega = 2\pi
\nu$, giving rise to a voltage $V = h\nu/2e =\phi_0 \nu$, where $\nu = (R_n/\phi_0)I$.  

For increasing values of the ratio $R/\ell$, the time-dependent behavior is more conveniently described in terms of Josephson-vortex motion using a quasistatic approach.  The applied current entering at the origin produces a uniform sheet-current density $\bm K_I= {\hat \rho} K_{I\rho} = \hat \rho I/2\pi R$ at the annulus.  The resulting Lorentz force\cite{Likharev86} $\bm F_L = -\hat \theta F_L = -\hat \theta K_{I\rho} \phi_0$ induces the Josephson vortex to rotate in a clockwise sense around the annulus.    When $R/\ell \gg 1$ ($\ell/R \ll 1$), the Josephson core becomes very compact and the  dissipation there becomes quite large.  As a consequence, for the same current $I$, the vortex speed $v = 2\pi R {\overline V}/\phi_0$ and the phase-slip frequency $\nu = {\overline V}/\phi_0$ become smaller than in the opposite limit $\ell/R \gg 1$, which has the effect of reducing the effective resistance of the weak link, $R_{eff} = {\overline V}/I$.  This behavior is similar to that in sandwich-type annular junctions, as discussed in Appendix \ref{sandwich}.

To show this quantitatively, we first note that the time dependence of all quantities calculated from the exact solution $\phi(\theta)$ in Eq.\ (\ref{phiN=1}) can be obtained to good approximation by replacing $\theta$ by $\theta + \omega t$.  The voltage measured directly across the weak link between $(\rho,\theta) = (R_-,\theta)$ and $(R_+,\theta)$ is $V(\theta,t) = (\hbar/2e)d\phi/dt = \phi_0 \nu \phi'(\theta+\omega t).$  The power delivered to the weak link by the external current source is therefore
\begin{equation}
P_{in} =  \frac{I \phi_0 \nu}{2 \pi} \int_{-\pi}^\pi \phi'(\theta+\omega t)d\theta = I {\overline V},
\label{Pin1}
\end{equation}
where ${\overline V}$,  the angular average of the voltage, is equal to the time-averaged voltage,  $<V> = h\nu/2e =\phi_0 \nu$. 
However, the power dissipated by the ohmic currents across the weak link\cite{Lebwohl67} is 
\begin{equation}
P_{out} = \int_{-\pi}^\pi [V^2(\theta+\omega t)/R_n]d\theta = ({\overline V}^2/R_n)\overline{\phi'^2},
\end{equation}
where the angular average of $\phi'^2$, obtained from Eq.\ (\ref{phiprimeN=1}), is
\begin{equation}
\overline{\phi'^2} = \frac{1}{2\pi} \int_{-\pi}^{\pi}[\phi'(\theta)]^2 d\theta = \sqrt{1+(R/\ell)^2}.
\label{avgthin}
\end{equation} 
Equating the input power $P_{in}$ to the dissipated power  $P_{out}$, we obtain the  effective resistance of the annular weak link,
\begin{equation}
R_{eff} = {\overline V}/I = R_n/\overline{\phi'^2}=R_n/\sqrt{1+(R/\ell)^2},
\label{Reff1}
\end{equation} 
and the corresponding phase-slip frequency, $\nu = (R_n/\phi_0)I/\sqrt{1+(R/\ell)^2}$.  

When $R/\ell \gg 1$, such that $\overline{\phi'^2} = R/\ell$ to good approximation, the Josephson core size ($\sim\!\!\ell$) becomes much smaller than the circumference of the weak link ($2\pi R$), and it is then appropriate to think of the Josephson vortex speed $v$ as being determined by a balance between the Lorentz force\cite{Likharev86} $F_L$ and a viscous drag force\cite{Lebwohl67} $\eta v$.  Equating the input power $P_{in} = F_L v$ to the dissipated power  $P_{out} = \eta v^2$, we obtain the viscous drag coefficient (units Ns/m),
\begin{equation}
\eta= \frac {\phi_0^2}{4\pi^2 R_n R \ell}.
\label{etathin}
\end{equation}
Note that $\eta$ is inversely proportional to the Josephson core size.  As discussed in Appendix \ref{sandwich}, this behavior of $\eta$ is similar to that in sandwich-type annular junctions, in which $\eta$ is inversely proportional to the Josephson penetration depth $\lambda_J$. 

The above calculations assume that the maximum value of the displacement current density across the weak link $(\epsilon_r \epsilon_0/d_N)(dV/dt)$ is much smaller than the maximum Josephson current density $j_c$.  This approximation is equivalent to the requirement that the vortex speed $v$ be much smaller than $\overline c$, where we find for $\ell/R \ll 1$,
\begin{equation}
\overline c = \Big(\frac{4\sqrt{3}\ell d_N}{9 \epsilon_r \Lambda d}\Big)^{1/2}c,
\end{equation}
where $\epsilon_r$ is the relative dielectric constant in the weak link and $c$ is the speed of light in vacuum.
Note that $\overline c$ is the analog of the Swihart velocity\cite{Swihart61} in long sandwich-type Josephson junctions.

\section{\label{exactallN}Exact solutions for arbitrary $N$}

When $N$ equally spaced Josephson vortices are trapped in the annular weak link ($N = 1,\;2,\;3,...$) with no Pearl vortex nearby and the current $I$ is zero, the Josephson vortex is stationary, and the gauge-invariant phase obeys
\begin{equation}
\frac{d\phi(\theta)}{d\theta}= N+ \frac{R}{2\pi \ell}\dashint_{-\pi}^\pi  \sin \phi(\theta')\cot \Big(\frac{\theta-\theta'}{2}\Big)d\theta'.
\label{phiprimeIntEqN}
\end{equation}
An exact solution of this equation, corresponding to one Josephson vortex centered at $\theta = 0$ and the others arranged around the annulus with  equal angular spacing $\Delta \theta = 2\pi/N$  is
\begin{equation}
\phi(\theta)= 2\tan^{-1}\Big[\frac{\tan(N\theta/2)}{\tan(N\theta_N/2)}\Big]+\pi
\label{phiarbN}
\end{equation}
for  $-\pi/N \le \theta \le \pi/N$.  For $\theta$ outside this interval in the positive (negative) $\theta$ direction, multiples of $2\pi$ must be added to (subtracted from) Eq.\ (\ref{phiarbN}) to make $\phi(\theta)$ continuous with the property that $\phi(\pi)-\phi(-\pi) = 2\pi N$. Also  
\begin{equation}
\tan(N\theta_N/2)=\sqrt{1+(R/N\ell)^2}-R/N\ell,
\label{tanthetaN}
\end{equation}
or, alternatively,  
\begin{equation}
\frac{R}{\ell}=N\frac{1-\tan^2(N\theta_N/2)}{2\tan(N\theta_N/2)}.
\label{RbyellN}
\end{equation}
Note that $\tan(N\theta_N/2) \to 1$ when $\ell \to \infty$, and $\theta_N \to \ell/R \to 0$ when $\ell \to 0.$

\begin{figure}
\includegraphics[width=8cm]{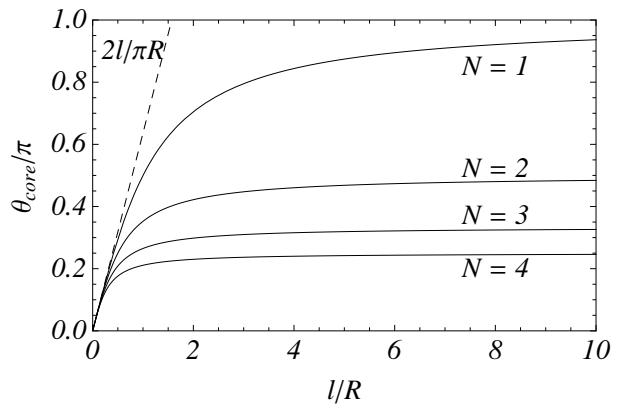}
\caption{%
Angular width of the Josephson core $\theta_{core}$ (in units of $\pi$) vs $\ell/R$ for $N = 1$, 2, 3, and 4 [Eqs.\ (\ref{thetacore}) and (\ref{thetaNcore})].  The dashed line shows the limiting behavior for $N\ell/R \ll 1$.}
\label{thetacoreplot}
\end{figure}

Equation (\ref{phiarbN}) yields 
\begin{equation}
\phi'(\theta) =  \frac{N \tan(N\theta_N/2)}{\tan^2(N\theta_N/2)\cos^2(N\theta/2)+\sin^2(N\theta/2)}.
\label{phiprimearbN}
\end{equation}
Note that $\phi'(0) = N\cot(N\theta_N/2) = \sqrt{N^2+(R/\ell)^2}+R/\ell$     and 
$\phi'(\pm \pi/N) = N\tan(N\theta_N/2) = \sqrt{N^2+(R/\ell)^2}-R/\ell$.
Equation (\ref{phiarbN}) also yields
\begin{equation}
\sin\phi(\theta)= - \frac{2 \tan(N\theta_N/2)\sin(N\theta/2)\cos   (N\theta/2)}{\tan^2(N\theta_N/2)\cos^2(N\theta/2)+\sin^2(N\theta/2)},
\label{sinphiarbN}
\end{equation}
which has a maximum (+1) and a minimum (-1) at $\theta=-\theta_N$ and  $\theta=+\theta_N $.  Defining the angular width $\theta_{core}$ of one of the Josephson cores as the range of $\theta$ values for which $\pi/2 \le \sin\phi(\theta) \le 3\pi/2$ (modulo $2\pi$), we find  for arbitrary $N$,
\begin{equation}
\theta_{core} = 2 \theta_N  = (4/N)\tan^{-1}[\sqrt{1+(R/N\ell)^2}-R/N\ell].
\label{thetaNcore}
\end{equation}
When $\ell/R \to \infty$, $\theta_{core} = \pi/N$, and when $\ell/R \ll 1$, $\theta_{core} \approx 2\ell/R.$ See Fig.\ \ref{thetacoreplot}.

As in the case $N=1$, the critical current of the annular weak link is zero for all $N$.  When a current $I$ is applied, the effective resistance can be calculated as in Eqs.\ (\ref{Pin1})-(\ref{Reff1}), except that for arbitrary $N$ we have  ${\overline V} =N\phi_0 \nu$, $P_{out} =  [(\phi_0 \nu)^2/R_n]\overline{\phi'^2},$ and 
\begin{equation}
\overline{\phi'^2} =  N^2\sqrt{1+(R/N\ell)^2},
\label{avgthinN}
\end{equation}
such that the effective resistance of the annular weak link is 
\begin{equation}
R_{eff} = {\overline V}/I = R_nN^2/\overline{\phi'^2}=R_n/\sqrt{1+(R/N\ell)^2},
\label{ReffN}
\end{equation} 

When $R/N\ell \gg 1$, such that $\overline{\phi'^2} = NR/\ell$ to good approximation, the Josephson core size ($\sim\!\!\ell$) becomes much smaller than the intervortex spacing ($2\pi R/N$).  In this case the effective resistance of the annular weak link containing $N$ Josephson vortices is $R_{eff} = NR_1$, where $R_1 = R_n\ell/R$ is the effective resistance when $N= 1$ in this limit [see Eq.\ (\ref{Reff1})].  It is also appropriate in this limit to think of the Josephson vortex speed $v$ as being determined by a balance between the Lorentz force\cite{Likharev86} $F_L$ and a viscous drag force\cite{Lebwohl67} $\eta v$.  Equating the input power per vortex $P_{in}/N = F_L v$ to the dissipated power  per vortex $P_{out}/N = \eta v^2$, we obtain exactly the same viscous drag coefficient as in Eq.\ (\ref{etathin}).

\section{\label{NearbyPearlVortex}Critical current affected by a nearby Pearl vortex}

We next consider the behavior when there is no flux quantum in the annular weak link ($N = 0$) but there is a Pearl vortex at $(x,y) = (\rho_v,0)$ either inside  the annulus ($\rho_v < R_-$) or outside ($\rho_v > R_+$).  For simplicity let us consider only the case for which $\ell/R$ is so large that we can ignore the effect of the Josephson currents on $d\phi(\theta)/d\theta$.  The equation determining the angular dependence of the gauge-invariant phase is then simply
 \begin{equation}
\frac{d\phi(\theta)}{d\theta} = P(\tilde \rho_v,\theta),
\label{phiprimePearl}
\end{equation}
where $\tilde \rho_v = \rho_v/R$ and $P(\tilde \rho_v,\theta)$ is given in Eq.\ (\ref{Pdefinition}).  Integration of Eq.\ (\ref{phiprimePearl}) yields the gauge-invariant phase difference,
\begin{equation}
\phi_v(\tilde \rho_v,\theta) = \theta-2\tan^{-1}[\gamma_v \tan(\theta/2)],
\end{equation}
where
\begin{equation}
\gamma_v=\Big|\frac{\tilde \rho_v +1}{\tilde \rho_v -1}\Big|
\end{equation}
and the constant of integration is chosen such that $\phi_v(\tilde \rho_v,\theta)=0$ at $\theta =0$, the point on the annulus that is closest to the Pearl vortex.

To calculate the critical current [see Sec.\ \ref{phase}], we note that the net supercurrent carried through the weak link is $I = I_{c0}\overline{\sin\phi}$.  Noting that $\phi(\theta) = \phi_v(\tilde\rho_v,\theta)+\beta$, where $\beta$ is a constant bias phase,  also is a solution of Eq.\ (\ref{phiprimePearl}), we obtain supercurrent-carrying solutions for which $I = I_{c0}\overline{\cos\phi_v }\sin\beta$.  The critical current is then given by the simple result,
\begin{equation}
I_c/I_{c0} = |\overline{\cos\phi_v}|=4\gamma_v/(1+\gamma_v   )^2
\label{Icgamma}
\end{equation}
or
\begin{eqnarray}
I_c/I_{c0} &=&1-\tilde\rho_v^2,\;0 \le \tilde\rho_v \le 1, \\
&=&1-\tilde\rho_v^{-2},\;\tilde\rho_v \ge 1.
\label{Icvoutside}
\end{eqnarray}
Note that $I_c = 0$ when $\tilde\rho_v = 1$, which corresponds to the case that the Pearl vortex has moved into the annular junction.  This is equivalent to the state $N = 1$ discussed in Sec.\ \ref{exactsolution}.

Equations (\ref{Icgamma})-(\ref{Icvoutside}) are valid only in the limit $R/\ell = 0$.  To calculate $I_c$ for finite values of $R/\ell$ would require solving Eq.\ (\ref{phiprimegeneral}) for $N=0$ at all $\rho_v$.  While this  equation can be solved perturbatively for small $R/\ell,$  the corrections to Eqs.\ (\ref{Icgamma})-(\ref{Icvoutside}) are second order in $R/\ell,$ such that this procedure yields only very small increases in the values of $I_c/I_{c0}$ for $0 < \tilde\rho_v < 1$ and $\tilde\rho_v > 1$.  How $I_c$ is affected for small values of $\ell/R$ (large $R/\ell$) remains unknown.

\section{\label{summary}Summary}

In this paper I have reported a detailed study of the properties of a Corbino-geometry annular weak link of radius $R$ in a superconducting thin film for which the Pearl length\cite{Pearl64} $\Lambda = 2\lambda^2/d$ is much larger than $R$.  I have considered separately the contributions due to  an integral number $N$ of flux quanta trapped in the weak link, a  Pearl vortex pinned nearby, and the Josephson current distribution across the weak link.   I derived two equivalent integral equations describing the gauge-invariant phase distribution $\phi(\theta)$ around the annulus, and I described how these integral equations can be transformed into each other.   I considered the case of $N = 1$ with no nearby Pearl vortex, first presenting an exact solution for $\phi(\theta)$ in the static case when $I = 0$, and then discussing the dynamic case for $I > 0$, when the Josephson vortex rotates around the annulus at constant angular velocity.  I then briefly discussed the case of an arbitrary number $N$ of equally spaced flux quanta trapped in the weak link, again presenting an exact solution for the static case when $I = 0$ and discussing the dynamic case when $I > 0$.  Finally, I calculated the critical current $I_c$ of the weak link as a function of the position of a nearby Pearl vortex and showed that $I_c = 0$ when the Pearl vortex falls into the weak link.

I mentioned in the introduction that thin-film annular weak links containing trapped vortices have been proposed\cite{Gaitan96,Gaitan01,Plerou01} as a place to test for the influence of the Berry phase on the vortex dynamics.  However, in this paper I have assumed that the vortex motion is determined only by the principle of conservation of energy: the vortex speed was obtained by setting  the power supplied to the weak link equal to the power dissipated via ohmic  currents.  I leave it to other authors to discover how this treatment may need to be modified to account for the influence of the Berry phase.

\begin{acknowledgments}
I thank J. E. Sadleir, R. H. Hadfield, M. G. Blamire, and V. G. Kogan for stimulating comments and helpful suggestions.
This work was supported by the U.S. Department of Energy, Office of Basic Energy Science, Division of Materials Sciences and Engineering. The research was performed at the Ames Laboratory, which is operated for the U.S. Department of Energy by Iowa State University under Contract No. DE-AC02-07CH11358.

\end{acknowledgments}
\appendix 

\section{$N$ flux quanta in a circular slot\label{Ninslot}}

The vector potential describing the magnetic flux density ${\bm B} = \nabla \times {\bm A}$ when there are $N$ flux quanta trapped in a narrow annular slot of radius $R$ in an otherwise thin film characterized by the Pearl length $\Lambda = 2 \lambda^2/d$ can be obtained using a procedure similar to that in Ref.\ \onlinecite{Pearl64} with the result ${\bm A} = \hat \theta A_\theta (\rho,z)$, where 
\begin{equation}
A_\theta (\rho,z)=\frac{N \phi_0}{2\pi}\int_0^\infty\frac{J_0(qR)J_1(q\rho)}{1+q\Lambda}e^{\mp qz}dq,
\end{equation}
and the upper (lower) sign holds when $z\ge 0$ ($z < 0$).  For $r=\sqrt{\rho^2 + z^2}  \gg \Lambda,$ ${\bm B} \approx \pm \hat r N\phi_0/2\pi r^2$, where $\hat  r = (\hat \rho \rho +\hat z z)/r$.  The sheet-current density is ${\bm K} = \hat \theta K_\theta(\rho)$, where
\begin{equation}
K_\theta(\rho)=\frac{N \phi_0}{\pi \mu_0 \Lambda}\Big[\frac{1}{\rho}\Theta(\frac{\rho}{R})-\int_0^\infty\frac{J_0(qR)J_1(q\rho)}{1+q\Lambda}dq\Big],
\end{equation}
and $\Theta(x)=1$ when $x>1$ and $\Theta(x)=0$ when $x < 1$.
  
\section{Sheet-current density\label{Kappendix}}

The Josephson-current-generated sheet-current density  can be evaluated analytically from the exact solution given in Eq.\ (\ref{phiN=1}) as follows.  The complex current density ${\cal K}(\zeta) = K_y(x,y)+i K_x(x,y) = d{\cal G}(\zeta)d\zeta$ can be obtained by differentiation of Eq.\ (\ref{G}).  The corresponding $\tilde {\cal K}(\zeta) ={\cal K}(\zeta) \zeta/\rho = K_\theta(\rho,\theta)+i K_\rho(\rho,\theta)$ [see Eq.\ (\ref{tildecalK})], where $\zeta = x + i y=\rho e^{i\theta}
$, $x = \rho \cos \theta$, $y = \rho \sin \theta$,  $\rho = \sqrt{x^2+y^2}$, and $\theta = \tan^{-1}(y/x)$, is then 
\begin{equation}
\tilde {\cal K}(\zeta) = \pm i \frac{K_c R }{2\pi\rho} \!\int_{-\pi}^\pi  \sin \phi(\theta')\Big(\frac{\zeta+\zeta'}{\zeta-\zeta'}\Big)d\theta',
\label{tildecalKintegral}
\end{equation} 
where $\sin\phi(\theta)$ is given by Eq.\ (\ref{sinphiN=1}), $\zeta' = R e^{i\theta'}$ and the upper (lower) sign holds when $\rho > R$ ($\rho < R$).  
Changing variables to $u = \tan[(\theta'-\theta)/2)$, $\tau = \tan(\theta/2)$, and $\tau_\ell = \tan(\theta_1/2)$ [Eq.\ (\ref{tantheta1})], and employing the definition
\begin{equation}
\alpha_\ell = \frac{1-\tau_\ell}{1+\tau_\ell} = \frac{R/\ell}{1+\sqrt{1+(R/\ell)^2}},
\end{equation}
reduces the apparent complexity of the resulting integral, whose evaluation yields for $\bar \rho = \rho/R > 1$,
\begin{eqnarray}
K_\theta(\rho,\theta)&=&\!\!-\frac{4K_c \tau_\ell[\alpha_\ell(\tau^2\!+\!1)+\bar \rho(\tau^2\!-\!1)]}{\bar \rho(1\!+\!\tau_\ell)^2[(\alpha_\ell\!-\!\bar\rho)^2\!+\!(\alpha_\ell\!+\!\bar\rho)^2\tau^2]}, \\
K_\rho(\rho,\theta)&=&\!\!-\frac{8 K_c \tau_\ell \tau}{(1\!+\!\tau_\ell)^2[(\alpha_\ell\!-\!\bar\rho)^2\!+\!(\alpha_\ell\!+\!\bar\rho)^2\tau^2]},
\end{eqnarray}
and for $\bar \rho = \rho/R < 1$,
\begin{eqnarray}
K_\theta(\rho,\theta)&=&\!\!+\frac{4K_c \tau_\ell[(\tau^2\!-\!1)+\alpha_\ell\bar \rho(\tau^2\!+\!1)]}{(1\!+\!\tau_\ell)^2[(1\!-\!\alpha_\ell\bar\rho)^2\!+\!(1\!+\!\alpha_\ell\bar\rho)^2\tau^2]}, \\
K_\rho(\rho,\theta)&=&\!\!-\frac{8 K_c \tau_\ell \tau}{(1\!+\!\tau_\ell)^2[(1\!-\!\alpha_\ell\bar\rho)^2\!+\!(1\!+\!\alpha_\ell\bar\rho)^2\tau^2]}. 
\end{eqnarray}

Since $\bm K= \hat x K_x + \hat y K_y = \hat \rho K_\rho + \hat \theta K_\theta$, where $\hat \rho = \hat x \cos \theta + \hat y \sin \theta$ and $\hat \theta = \hat y \cos \theta - \hat x \sin \theta$, the above results also yield $K_x$ and $K_y$  via $K_x = K_\rho \cos \theta - K_\theta \sin \theta$ and $K_y = K_\rho \sin \theta + K_\theta \cos \theta$.  

In the limit as $\bar \rho \to 1$ ($\rho \to R_+$ or $R_-$), the above results reduce to
 \begin{eqnarray}
K_\theta(R_{\pm},\theta)&=&\!\!\mp\frac{2K_c \tau_\ell(\tau^2\!-\!\tau_\ell)}{(1\!+\!\tau_\ell)(\tau_\ell^2+\tau^2)}, \\
K_\rho(\rho,\theta)&=&\!\!-\frac{2 K_c \tau_\ell \tau}{(\tau_\ell^2+\tau^2)}. 
\end{eqnarray}
However, $K_\theta(R_\pm,\theta)$ can be derived more simply from Eq.\ (\ref{phiprimegeneral}) using $N = 1$, $P = 0$, and Eqs.\ (\ref{Kthetaplusminus}), (\ref{Rbyell}), and (\ref{phiprimeN=1}), while $K_\rho(R,\theta)$ can be derived from $K_\rho(R,\theta)= K_c \sin\phi(\theta)$ and  Eq.\ (\ref{sinphiN=1}).

When $R/\ell = 0$, $\alpha_\ell = 0$, and $\tau_\ell = 1$, the current pattern is dipole-like.  For $\bar \rho > 1,$ $K_\theta (\rho,\theta) = K_c\cos \theta/\bar \rho^2$
and $K_\rho (\rho,\theta) = -K_c\sin \theta/\bar \rho^2$, while for $\bar \rho < 1,$ $K_\theta (\rho,\theta) = -K_c\cos \theta$
and $K_\rho (\rho,\theta) = -K_c\sin \theta.$

For other values of $R/\ell$, the current pattern is dipole-like only at distances $\bar \rho \gg \alpha_\ell$, where, to good approximation,
\begin{eqnarray}
K_\theta(\rho,\theta)&=&\!\!\frac{4K_c \tau_\ell\cos\theta}{(1\!+\!\tau_\ell)^2\bar\rho^2}, \\
K_\rho(\rho,\theta)&=&\!\!-\frac{4K_c \tau_\ell\sin\theta}{(1\!+\!\tau_\ell)^2\bar\rho^2}. 
\end{eqnarray}

\section{Comparison with sandwich-type annular junctions\label{sandwich}}

Numerous experimental studies have been carried out in annular Josephson junctions, with some of these having the so-called Lyngby geometry.\cite{Davidson85}  These junctions can be thought of as long ring-shaped Josephson junctions sandwiched between a pair of superconducting washers.  In such junctions, there is only one length scale, the Josephson penetration depth\cite{Josephson65} $\lambda_J = (\phi_0/2\pi \mu_0 j_cd_{eff} )^{1/2}$, characterizing the spatial variation of both the magnetic field and the non-linear core of a Josephson vortex as a function of the coordinate $x$ along the length of the junction.  Here $j_c$ is the maximum Josephson supercurrent, $d_{eff} = d_i + 2 \lambda$, where $d_i$ is the insulating barrier thickness and $\lambda$ is the London penetration depth, and it is assumed that $\lambda_J \gg d_{eff}$.   Starting with solutions of the sine-Gordon equation to describe the gauge-invariant phase distribution associated with a periodic Josephson-vortex array of period $L$, Lebwohl and Stephen\cite{Lebwohl67} discussed the Lorentz-force-induced motion of the array and calculated the resulting viscous drag coefficient per unit length of vortex $\eta$.  

\begin{figure}
\includegraphics[width=8cm]{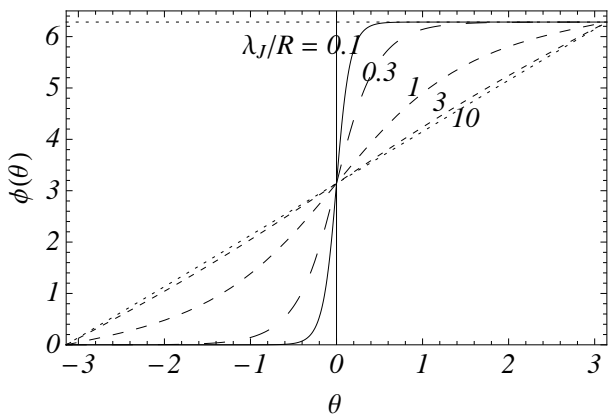}
\caption{%
Plots of the solution $\phi(\theta)$ [Eqs.\ (\ref{phisandwich})-(\ref{phisandwich3})] vs $\theta$ for a Josephson vortex ($N = 1$) in a sandwich-type annular weak link for $\lambda_J/R =$ 0.1, 0.3, 1, 3, and 10.}
\label{phisandwichplot}
\end{figure}

The gauge-invariant phase distribution for a sandwich-type annular weak link of radius $R$ and width $W$, where $d_{eff} \ll W \ll 2\pi R$, containing a single Josephson vortex ($N = 1$) can be obtained from Ref.\ \onlinecite{Lebwohl67} by simply replacing $L$ by $2\pi R$.  The solution analogous to Eq.\ (\ref{phiN=1}) is 
\begin{equation}
\phi(\theta) =  2 \sin^{-1}\Big[{\rm sn}\Big(\frac{R\theta}{k\lambda_J}\Big|k^2\Big)\Big]+\pi,
\label{phisandwich}
\end{equation}
where $kK(k)= \pi R/\lambda_J$, $K(k)$ is the complete elliptic integral of the first kind\cite{Abramowitz67,Gradshteyn00} of modulus $k$ and parameter $k^2$, and sn(u$|$m)is the Jacobian elliptic function of parameter $m$.  In the limits of small and large $R/\lambda_J$, Eq.\ (\ref{phisandwich}) reduces to 
\begin{eqnarray}
\phi(\theta) &=&  \theta+\pi,\;R/\lambda_J \ll 1,\\
&=&2\sin^{-1}[\tanh(R\theta/\lambda_J)]+\pi, \;R/\lambda_J \gg 1.
\label{phisandwich3}
\end{eqnarray}
The angular dependence of $\phi(\theta)$ in a sandwich-type annular weak link is displayed in Fig.\ \ref{phisandwichplot}.

The critical current of a sandwich-type annular junction containing a single Josephson vortex is zero, and for small currents it is a good approximation to assume that the phase distribution of Eq.\ (\ref{phisandwich}) rotates around the annulus with a speed $v = R\omega = 2\pi R \nu$.   The effective resistance of the junction, calculated as in Sec.\  \ref{exactsolution}, is
\begin{equation}
R_{eff} = {\overline V}/I = R_n/\overline{\phi'^2},
\label{Reffsandwich}
\end{equation} 
where $R_n$ is the normal-state resistance of the annular junction,  $\overline{\phi'^2}$ is the angular average of $[d\phi(\theta)/d\theta]^2$,  
\begin{eqnarray}
\overline{\phi'^2}&=&\frac{4R E(k)}{\pi \lambda_J k},\\
&=& 1,\;R/\lambda_J \to 0,\\
&=& \frac{4R}{\pi \lambda_J},\;R/\lambda_J \gg 1,
\label{avgsandwich}
\end{eqnarray}
and $E(k)$ is the complete elliptic integral of the second kind.\cite{Abramowitz67,Gradshteyn00}

In the limit $\lambda_J \ll R$, for which the Josephson core size ($\sim\!\lambda_J$) is much smaller than the circumference of the annulus ($2\pi R$), it is appropriate to think of the effective resistance as arising from a balance between the Lorentz force per unit length of vortex and a viscous drag force per unit length.  In this limit, the viscous drag coefficient per unit length (units Ns/m$^2$) is\cite{Lebwohl67}
\begin{equation}
\eta=\frac{\phi_0^2}{\pi^2 R_n R W \lambda_J},
\label{etasandwich}
\end{equation}
where  $W$ is the width of the annular junction.  Note that $\eta$ is inversely proportional to the Josephson core size.

\end{document}